# New security and control protocol for VoIP based on steganography and digital watermarking

Wojciech Mazurczyk[1] and Zbigniew Kotulski[1,2]

[1] Warsaw University of Technology, Faculty of Electronics and Information Technology, Institute of Telecommunications
{W.Mazurczyk Z.Kotulski}@tele.pw.edu.pl

[2] Polish Academy of Sciences, Institute of Fundamental Technological Research
zkotulsk@ippt.gov.pl

**Abstract**
In this paper new security and control protocol for Voice over Internet Protocol (VoIP) service is presented. It is the alternative for the IETF's (Internet Engineering Task Force) RTCP (Real-Time Control Protocol) for real-time application's traffic. Additionally this solution offers authentication and integrity, it is capable of exchanging and verifying QoS and security parameters. It is based on digital watermarking and steganography that is why it does not consume additional bandwidth and the data transmitted is inseparably bound to the voice content.

**Keywords:** IP Telephony / VoIP security, digital watermarking, steganography

## 1. Introduction

Nowadays two the most important fields in which Voice over Internet Protocol (VoIP) is lacking are providing certain Quality of Service (QoS) parameters and security considerations [3]. In this paper we consider new protocol that covers both those fields simultaneously. It provides information that is vital to control the network conditions and to verify authentication of the source and data integrity.

In TCP/IP networks VoIP, which is a real-time service, uses RTP (Real-Time Protocol) with UDP (User Datagram Protocol) for transport of digital streams. Currently there is one control protocol for RTP and it is RTCP (Real-Time Control Protocol) [5]. It is designed to monitor the Quality of Service (data delivery) and to convey information about the participants in an on-going session. RTCP operates mainly on exchanging two special reports called: Receiver Report (RR) and Sender Report (SR). Parameters that are enclosed in those reports can be used to estimate the network status.

We propose a new protocol that uses two techniques: digital watermarking and network steganography to achieve analogous functionality like RTCP but moreover offers additional advantages. The most important one is that it includes also security verification of the transmission source and the content send (authentication and integrity). This solution does not consume transmission bandwidth, because the control bits (a header of the new protocol) are transmitted in a covert (steganographic) channel and data (QoS and the security parameters) is inseparably bound to voice content as a watermark.

The paper is organized as follows. In Section 2 both techniques, digital watermarking and steganography, are described. Next, we give details about proposed solution in Section 3. Finally, we end with conclusions in Section 4.

## 2. Steganography and Digital Watermarking
Steganography and Digital Watermarking are Information Hiding subdisciplines. The general difference between those two techniques is that steganography's aim is to keep the existence of the information secret and in watermarking making it imperceptible.

### 2.1. Steganography: a covert channel
Steganography is a process of hiding secret data inside other, normally transmitted data. Usually it means hiding a secret message within an ordinary message and the extraction of it at its destination. In ideal situation, anyone scanning data will fail to know it contains covert data. In modern digital steganography, data is inserted into redundant (provided but often unneeded) data, e.g. fields in communication protocols, graphic image, etc. TCP/IP steganography utilize the fact that few headers in packet are changed during transit. We will exploit here a covert channel, which is a method of communication that is not a part of an actual computer system design, but can be used to transfer information to users or system processes that normally would not be allowed access to the information. In TCP/IP stack, there is a number of methods available, whereby covert channels can be established and data can be exchanged secretly between hosts. An analysis of the headers of typical TCP/IP protocols e.g. IP, UDP, TCP, HTTP, ICMP results in fields that are either unused or optional. This reveals many possibilities where data can be stored and transmitted. As described in [7] IP header posses few fields that are available to be used as a covert channel. Those fields are marked in Figure 1 with italics. The total capacity of those fields exceeds 60 bits per packet. And there are UDP and RTP protocols fields left to be used.

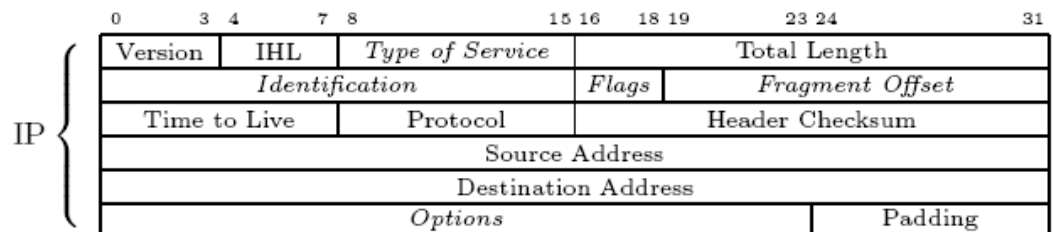
Figure 1 IP header with marked fields (italics) available for steganography (source: [7])

For VoIP and our solution we will exploit unused/optional fields in IP/UDP/RTP packets because those protocols are used in almost all IP telephony implementations. In [7], [8] and [9] methods of hidden transmission are presented. We do not limit this solution to using only IP/UDP/RTP protocol. Lower layers of TCP/IP stack also offer steganography possibilities like for example stated in [10]. Furthermore we can distribute those control bits among those fields in a predetermined fashion (this pattern can be exchanged during a signalling phase of conversation). In those chosen fields we will transmit only header (control bits) of our protocol with the use of steganography technique. Header consists of 6 bits per packet, so such a type of transmission is potentially hard to discover. The details will be described in Section 3.

### 2.2. Watermarking: the imperceptible information
Digital watermarking is a multidisciplinary methodology widely developed in the last decade. It covers a large field of various aspects, from cryptography to signal processing, and is generally used for marking the digital data (images, video, audio or text). There are several applications for digital watermarks, described in [1] and [2], that include: **Fingerprinting** (embedding a distinct watermark into every copy of the author's data), **Annotation Watermark/Content labelling** (embedding information, which describes the digital work that can be later extracted), **Usage control/Copy control** (authors can insert a watermark that indicates the number of copies permitted for each user). However the most important applications for our purposes are: the possibility of embedding the **authentication and integrity watermark** and exchanging additional information (for the controlling RTP packets purposes).

The watermark that will be used in, proposed here, authentication and integrity solution must possess certain parameters, like: robustness, security, transparency, complexity, capacity, verification and invertibility. Those parameters are described in [1] and [2]. Their optimization for real-time audio system is crucial. They are often mutually competitive, however there is always a compromise necessary. That is why the embedded watermark, that we will use, must be characterized by **high robustness**, **high security** and must be **non-perceptual**. Not every watermarking technique is applicable for our solution. IP Telephony is the demanding, real-time service. That is why we need the watermarking schemes that really work for the real-time conversations. Such algorithms are described, e.g. in [2], [3] and [4].

Generally, audio watermarking algorithm is based on two functions: **embedding** of the watermark into voice and its **extraction**. As soon as the conversation begins, certain information is embed into the voice samples and sent through the communication channel. Then, the watermark is extracted from those samples before they reach the callee and the information retrieved is verified. If the watermark's data sent is correct, the conversation can be continued.

Most digital watermarking algorithms for the real-time communication are designed to survive typical non-malicious operations like: low bit rate audio compression, codec changes, DA/AD conversion or packet loss. For example, in [2] the watermarking scheme developed at the Fraunhofer IPSI (Institut Integrierte Publikations und Informationssysteme) and the Fraunhofer IIS (Institut Integrierte Schaltungen) were tested for different compression methods. Those results revealed that the large simultaneous capacity and robustness depend on the scale of the codec compression. When the compression rate is high (1:53), the watermark is robust only when we embed about 1 bit/s. With a lower compression rate we can obtain about 30 bit/s, whereas the highest data rate was 48 bit/s with good robust, transparent and complexity parameters. For the monophonic audio signal, which is a default type for IP Telephony the watermark embedding algorithm appeared around 14 times faster and the watermark detector almost 6 times faster than the real-time.

The next important thing for this scheme is how much information we can embed into the original voice data. This will influence the speed of the authentication and integrity process throughout the conversation. This parameter, in our solution, is expected to be high but it is not crucial. With low compression rates, we propose to add a pre-conversation stage. In this stage there will be few seconds of the RTP packets exchange without the conversation. It will delay the setup of the call but then, during the conversation, the time of verification will be shorter. However, the lowest payload watermarks (about 1 bit/s) cannot be accepted in our scheme because, in this case, the conversation would have to last enormously long to work correctly.

### 3. New control and security protocol

The most important security services to secure IP Telephony system are: **authentication**, **integrity** and **confidentiality**. The first two can be provided with the use of our protocol. The third should be guaranteed in a different manner, e.g., with the use of the security mechanisms from a classical security model (the cryptographic mechanisms).

As we described earlier we will use two information hiding techniques: steganography to create covert channel that will be used to transmit header (control bits) and digital watermarking to bound the parameters of the protocol to voice send into the network (watermark). We assume that our solution is for using in IP protocol version 4 networks [14]. The protocol we are proposing here should posses PDU (Protocol Data Unit), of which size must be kept to minimum. It is important because as we said in the Section 2 the capacity capability of watermarks is limited if we want also watermark other parameters like robustness or security. Every PDU consists of header (control bits) and a certain number of data bits that are embedded into sender/receiver voice. Because the capacity of the watermark depends greatly on the codec's compression rate that is used, so it is possible that the lot of parameters can be distributed into a number of packets. The size (number of bits) of each parameter that will be transmitted with our protocol should be low. For all parameters it should not exceed 32 bits. This value is taken from RTCP protocol size of the parameters. Only one parameter (NTP timestamp) is greater than given value. Limited size of every parameter results in shorter time for the parameter to be transmitted and verified. However we do not dictate this value. It should depend on network bandwidth, status and codec's compression rate.

### 3.1. Protocol data unit description

The PDU consists of two parts: the header (control bits) and the watermark data. The header/control fields are transmitted in a covert channel in unused/optional fields of IP/UDP/RTP protocol's headers. Actual value of the parameter is embedded into voice as a watermark.

The header (control bits) are organized in fields as shown in the Table 1:

| Type of field | Number of bits | Function |
|---|---|---|
| **P** (Parameter) | 4 | Describes parameter that is transmitted in the watermark |
| **S** (Side) | 1 | Describes the side of the communication (1 - sender, 0 - receiver report) |
| **C** (Continuity) | 1 | Describes if a packet contains the beginning or continuation of the parameter indicated in the field P (1 – beginning of new parameter, 0 – continuation of the last parameter) |

Table 1 Header fields and their function

Exemplary values of the field P are shown below (analogous parameters like in RTCP [5]):
    0001 – authentication or integrity parameter **(**32 bits**)**
    0010 – parameter: LSR – Last sender report (32 bits**)**
    0011 – parameter: DLSR – Delay of last sender report (32 bits)
    0100 – parameter: Interarrival jitter (32 bits)
    0101 – parameter: Extend highest sequence number received (32 bits)
    0111 – parameter: Cumulative number of packet lost (24 bits)
    1000 – parameter: Fraction lost (8 bits)
    1001 – parameter: Sender's packet count (32 bits)
    1011 – parameter: NTP timestamp (64 bits)
    1010 – parameter: RTP timestamp (32 bits)
    …

Moreover, the PDU can have one of two payload types: **security** or **informational**. Security payload means that PDU contains certain authentication and/or integrity information that should be verified after its extraction. Two kinds of security payloads are available, first is used to provide authentication and integrity of the voice and its source. Second's role is to authenticate protocol parameters that were send earlier (both security and informational). Details about security payload and cryptographic operations in this protocol will be covered in Section 3.2.

Another payload type is informational. Each PDU carries one of the parameters that are used to monitor the quality of service and the network conditions.

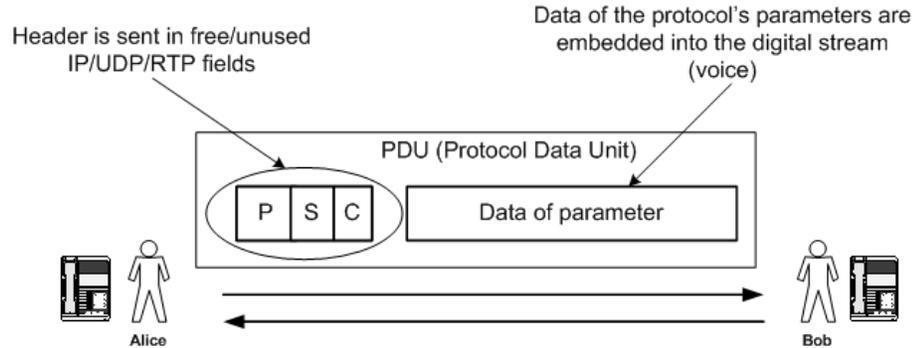

Fig. 2. General protocol operation

Usually in one IP/UDP/RTP packet there is about 20-30 milliseconds of voice, which is about 20-30 bytes, depending on type of codec used. Lets say that we are able to embed in average about 10 bits/s of watermark into the voice stream. With that assumption we must send about 3-4 packets to achieve those 10 bits. In this protocol we set parameter's value to 32 bits, so this parameter will be transmitted in about 9-12 packets in more than 3 seconds of the voice. In the example scenario in Fig. 3, we see how the exemplary parameter: Interarrival jitter (32 bits) is transmitted for assumption: 10bits/packet.

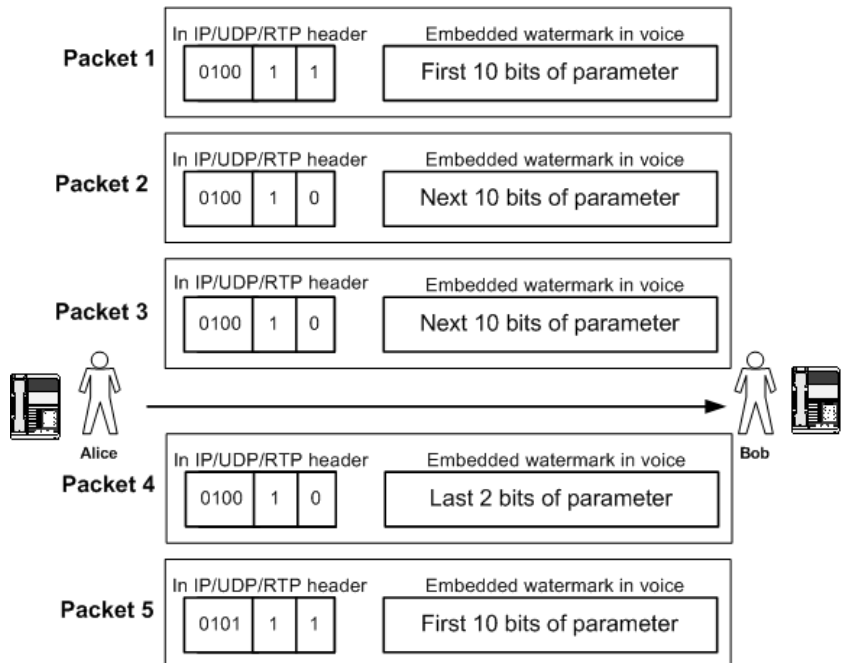

Fig. 3. Example of transmission of the Interarrival jitter parameter

As we can see in Fig. 3 parameter characterized by code 0100 (Interarrival jitter) was sent in four IP/UDP/RTP packets. In the first packet both fields S and C was set to 1. In the next packet field C changed its value to 0 because it is a continuation of the parameter's data that was sent in last packet. At the destination there must be a buffer to extract all data from each packet. After transmitting all packets for one parameter data is available to be used (for QoS monitoring) or to be verified (for security reasons).

**3.2. Authentication and integrity parameter calculation and security payload**
Authentication and integrity calculation will be performed similarly as described in [6] but with watermark specific considerations.
In Section 3.1 we mentioned that two security payloads are available:
- one is used to provide authentication and integrity of the voice and its source
- second is to authenticate protocol parameters (both security and informational) that were send earlier

First security parameter is a combination of user global identification and features that were extracted from the voice stream. It is expected that this parameter will have 32 bits. So if the concatenation of those two values exceeds this number of bits, there will be a hash function (marked as H) performed. Then only predetermined bits will be transmitted as a security parameter.

Second security payload is a special parameter that will be used to provide greater security of the whole digital stream and transmission. The general idea of its calculation is presented in Figure 4.

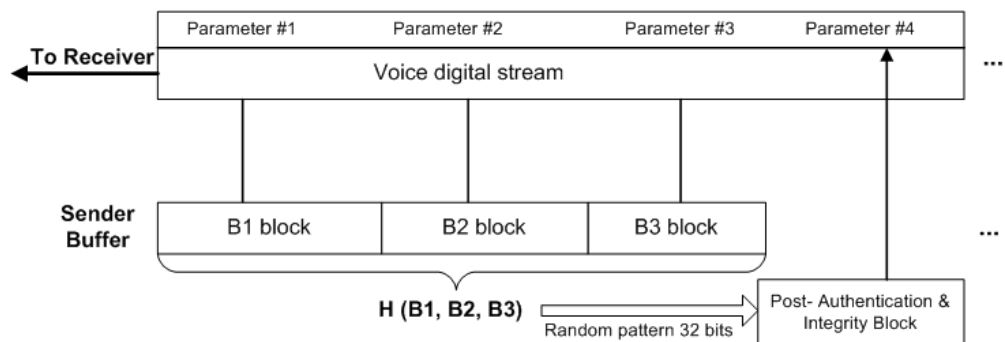

Fig. 4. Example of authentication and integrity mechanism for transmitted parameters

First, we must emphasize that during a conversation (RTP packets flow) there will be constant two-way exchanging of certain sequence of parameters. Those parameters can be susceptible to e.g. modifications or other attacks. To prevent this situation every n-th packet is used to authenticate and provide integrity of n-1 parameters that were transmitted earlier.
For the situation in Fig. 4. n=4. Three parameters that can contain informational or first kind of security payload are stored in sender buffer. After they all are placed there (B1, B2 and B3 blocks) the hash function (H) can be calculated, if the result value is too long. Because we assumed certain parameter length that is why we have to choose only 32 bits from the hash to be transmitted. For every conversation this pattern, in which bits are chosen, should be changed and its determination should be set and send in a signalling phase of the connection.

**3.3. Level of Trust (LoT) mechanism**
However, we can still imagine a situation, in which the attacker disrupt transmission of the header/controls bits. In this situation the receiver is unable to retrieve any parameters that were transmitted by the sender. Here we will describe a mechanism that will prevent such a situation. Both parties of the conversation will update special parameter named LoT (Level of Trust), during a conversation. If a parameter (security or informational) is received and verified, LoT value increases. In any other situation its value decreases. Additionally parameters that are exchanged during conversation influence LoT value differently. Informational parameters (QoS) add/subtract to LoT's value 1, first kind security parameters 2 and the second kind security parameter 5. If A sends to B a parameter, the algorithm of handling the LoT parameter (on B side) works, as described below in a pseudo-code:

```
START /* CL – Critical Level, LoT – Level of Trust, T – timer */
CL = a; LoTA = x; TA = 0; /* Initiating values */
StartTimer(TA);
FOR (i = 0; i++; i< End of Transmission)  /* i – Time slot */
  {
  IF (ParameterA correct) THEN
    {
    LoTA + {1 or 2 or 5}; *
    ResetTimer(TA);
    }
  ELSE (LoTA – {1 or 2 or 5}); *
 IF (LoTA <= CL) OR (TA > k) THEN STOP; (1)
 IF (LoTA = a*x) THEN LoT = x; (2)
}

* value depends on the type of parameter (QoS, security)
```

As we can see, the breakage of the call (or notification to the calling parties) will take place if the value of the LoT parameter is equal or below the given threshold (CL value) or if the timer TA expires (1). The LoT value changes during the conversation time. If every signalling message is successfully verified, the LoT value rises. To prevent its increase from reaching the infinity, we lower it, as soon as it reaches the value of the critical level multiplied by the start value of LoT (2).

This way of decreasing the LoT value has one serious disadvantage: it allows an attacker to wait until LoT= (a*x)-1. But we must assume that he is able to posses information about its value and then safely spoof ((a*x) - 1 – (CL + 1)) audio packets without LoT's falling below the threshold (CL). To prevent it, one must choose the initiating values (a and x) carefully. Their values should depend on network's parameters: the packet loss and possible delays. If the network does not suffer heavily from the packet loss, those values must be low. In the other case, they must be set to a higher level. For example, the network administrator or service provider can circumscribe those parameters for a certain network/user.

**4. Conclusions**
New security and control protocol for VoIP service was presented. It uses two information hiding techniques: steganography to create covert channel in which the header (control bits) are passed and digital watermarking to transmit the actual data (parameter's value) in voice stream. The most important advantages of our solution are no consuming of available bandwidth, providing security, parameters to monitor QoS and network status in one protocol. What we want to emphasize is that the process of sending information for this protocol is continuous in time and although the bit rate per second offered by watermarking is usually not very high, when we consider a whole conversation we see that we are able to exchange quite a amount of data.

The variety of different kind of parameters that can be used in our solution is not limited to security/monitoring status of the network ones. That is why this protocol can be freely extended to other data e.g. to support other and more detailed statistics as described in [11].